\documentclass[twocolumn,pre,floatfix]{revtex4}

\usepackage{graphicx}
\usepackage{epsfig}
\usepackage{rotating}
\usepackage{amsmath}
\usepackage{amsfonts}
\usepackage{amssymb}
\usepackage{enumerate}
\usepackage{longtable}
\usepackage{dcolumn}% Align table columns on decimal point
\usepackage{bm}
\newcommand{\chiSG}{\chi_{_{SG}}}
\newcommand{\av}{_{\mathrm{av}}}
\begin{document}

\title{Stability of the quantum Sherrington-Kirkpatrick spin glass model}

\author{A. P. Young}
\affiliation{University of California Santa Cruz, CA 95064, USA}

\date{\rm\today}

\begin{abstract}
We study in detail the quantum Sherrington-Kirkpatrick (SK) model, i.e.~the
infinite-range Ising spin glass in a transverse field, by solving numerically
the effective one-dimensional model that the quantum SK model can be mapped to
in the thermodynamic limit.  We find that the replica symmetric (RS) solution is
unstable down to zero temperature, in contrast to some previous claims, and so
there is not only a line of transitions in the (longitudinal)
field-temperature plane (the de Almeida-Thouless, AT, line) where replica
symmetry is broken,
but also a quantum de Almeida-Thouless (QuAT) line in the
transverse field-longitudinal field plane at $T = 0$. If the QuAT line also
occurs in models with short-range interactions its presence might affect the
performance of quantum annealers when solving spin glass-type problems with a
bias (i.e.~magnetic field). 
\end{abstract}

%\pacs{74.70.-b,75.10.Jm,75.40.Gb,75.30.Ds}

\maketitle

\section{Introduction}
\label{sec:intro}

Recently there has been a resurgence of interest in quantum spin glasses.
This is
motivated by the possibility that quantum annealing
(QA)~\cite{kadawoki:98,farhi_long:01} might be an
effective way to solve optimization problems. There have been both experiments
on real hardware made by D-Wave with up to around 2000 qubits~\cite{johnson:11}
and simulations both on special models~\cite{hen:11,farhi:12}
and on problems which have can be embedded naturally 
onto the D-Wave machine~\cite{ronnow:14,zhu:16}.

Bottlenecks in QA occur where the gap between the ground state and the first
excited state becomes very small. One situation where this occurs is at a
quantum phase transition, so it is useful to locate and characterize quantum
phase transitions in models that are commonly used for QA, which are
generally spin glasses. 

One of the most striking predictions of the mean field theory of spin
glasses~\cite{edwards:75,sherrington:75,binder:86} is the existence of a line
of transitions in the magnetic-field temperature plane first found by de
Almeida and Thouless (AT)~\cite{almeida:78}.
%This transition is
%surprising since it occurs without
%the breaking of any ``obvious" symmetry, and instead marks the onset of
%``replica symmetry breaking'' (RSB).
The solution of the mean-field,
infinite-range,
Sherrington-Kirkpatrick (SK)~\cite{sherrington:75} model in the RSB phase
below the AT line is complicated and 
was obtained, in a tour-de-force,
by Parisi~\cite{parisi:80,parisi:83}.

The simplest approach to make a classical Ising model quantum is to add a
transverse field $h^T$.  
Here we investigate phase transitions in the quantum
SK model including both a transverse field and a longitudinal field $h$. The
Hamiltonian is~\cite{rand_hT}
\begin{equation}
\mathcal{H} = -\sum_{\langle i, j \rangle} J_{ij} \sigma^z_i \sigma^z_j - 
\sum_{i=1}^N h_i \sigma^z_i - h^T \sum_{i=1}^N \sigma^x_i \, ,
\label{ham}
\end{equation}
where the $\{\sigma^x_i, \sigma^z_i\}, (i = 1, \cdots, N)$ are Pauli spin
matrices, the interactions $J_{ij}$, which are between \textit{all} $N(N-1)/2$ pairs
of sites, have a Gaussian distribution with mean and variance given by
\begin{equation}
[J_{ij}]\av = 0, \qquad [J^2_{ij}]\av = J^2 / N\, ,
\end{equation}
the longitudinal fields $h_i$ have a Gaussian distribution with mean zero and
standard deviation $h$, and the transverse field $h^T$ is taken, for
simplicity, to be the same on each site. Here the notation $[\cdots]\av$
indicates an average over the quenched disorder.

\begin{figure}
\begin{center}
\includegraphics[width=0.8\columnwidth]{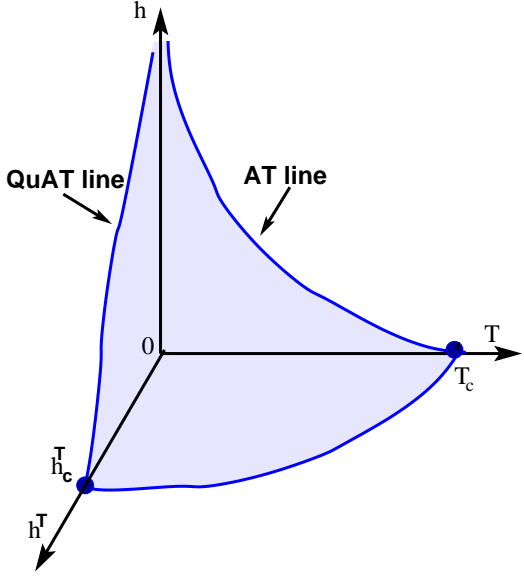}
\caption{\label{fig:quatline} 
A sketch of the proposed phase diagram of the quantum SK model in terms of
temperature $T$, the standard deviation of the (random) longitudinal field $h$, and
the transverse field $h^T$. The phase diagram for a \textit{uniform} longitudinal field $h$
would look the same. The replica symmetric solution is unstable everywhere
below the surface indicated. For $h^T = 0$ the boundary where replica symmetry
breaking occurs is called the AT line~\cite{almeida:78}. For $T=0$, we call
the boundary in the $h$-$h^T$ plane the quantum AT (QuAT) line. The phase
boundary in the $T$-$h^T$ plane is drawn \textit{precisely} in
Fig.~\ref{fig:Tc_hT}.
}
\end{center}
\end{figure}

For $h^T = 0$, de Almeida and Thouless~\cite{almeida:78} showed that one must
have replica symmetry breaking (RSB) below a line in the $h$-$T$ plane, see
Fig.~\ref{fig:quatline}. For $h=0$, the phase boundary in the $h^T$-$T$ plane 
has been extensively
studied~\cite{yamamoto:87,ray:89,lai:90,goldschmidt:90,buttner:90,read:95,mukherjee:15,mukherjee:17}.
The question
of whether RSB occurs in this plane all the way to $T=0$ has been
controversial. For example, Refs. ~\cite{ray:89} and \cite{mukherjee:15,mukherjee:17} argue that
there is a \textit{low-temperature} region with replica symmetry, while
Ref.~\cite{thirumalai:89} claims that there is region \textit{near the spin glass phase
boundary} with replica symmetry.  By contrast, 
Refs.~\cite{read:95,goldschmidt:90,buttner:90} argue
that \textit{replica symmetry is broken all the way
down to} $T = 0$.
%though in the latter work the situation seems to be marginal for $T$ strictly
%zero. 

Here we investigate the stability of the replica symmetric (RS) solution in
the $h^T$-$T$ plane finding that it is unstable all the way down to $T = 0$.
As a result there must be a quantum AT (QuAT) line in the $h$-$h^T$ plane at
$T=0$ which we investigate for small longitudinal fields $h$.
%We are not aware
%hat the QuAT line has been explicitly discussed in the literature before. 
If
it also occurs in those spin glass models which are used in QA studies, it
could affect the performance of QA on spin glass problems in a longitudinal
field.

The plan of this paper is as follows. In Sec.~\ref{sec:oned} we describe the
effective one-dimensional model whose solution gives the behavior of the
quantum SK model
in the thermodynamic limit. The results of our numerical simulations of this model are
described in Sec.~\ref{sec:results}, while the conclusions are summarized in
Sec.~\ref{sec:conclusions}.
%Some material of a more technical nature are
%discussed in the appendices.

\section{Reduction to an effective on-dimensional model}
\label{sec:oned}

It is, by now, standard, 
\cite{federov:86,buttner:90,ye:93,huse:93,read:95}
to reduce the quantum SK model in the
replica-symmetric (RS) phase to an effective, non-disordered,
long-range, one-dimensional
model in which the dimension corresponds to imaginary time $\tau$, running from $0$
to $\beta$, the inverse temperature. The interactions in this model have to be
determined self-consistently.  For our purposes it will be convenient to
discretize imaginary time into $M$ time slices, labeled by $l$, each of width
$\Delta\tau = \beta/M$.

We consider
first the case of zero longitudinal field. One finds that the free energy per
site $f$ is given by
\begin{multline}
-\beta f = 
-{1 \over 4} \beta\Delta \tau J^2 \sum_{\Delta l = 1}^M r(\Delta l)^2
+{1 \over 4} (\beta J)^2 q^2 \\
+ {1 \over \sqrt{2 \pi}} \int_{-\infty}^\infty dz \, e^{-z^2/2} \,
\ln \mathrm{Tr}\, e^{-\overline{H}(z)} \, ,
\label{fe}
\end{multline}
where $\overline{H}(z)$,
the effective Hamiltonian of the one-dimensional problem, is given by
\begin{multline}
\overline{H}(z) = -(\Delta\tau J)^2 
\sum_{\langle l_1, l_2\rangle} \left\{\,r(|l_1 - l_2|) - q\, \right\}
S_{l_1} S_{l_2} \\
- K^\tau \sum_l S_l S_{l+1} -
(\Delta \tau J) q^{1/2} z \sum_l S_l \, ,
\label{olineH}
\end{multline}
where $S_l$, the spin at time slice $l$, takes values $\pm 1$.
The variable $q$ is the spin glass order parameter defined by
\begin{equation}
q = [\, \langle \sigma_i^z \rangle^2\, ]\av \, ,
\end{equation}
and $K_\tau$, the nearest-neighbor coupling along the $\tau$ direction, is
given by
\begin{equation}
e^{-2 K^\tau} = \tanh(h^T \Delta\tau) \, .
\label{Ktau}
\end{equation}
The long-range interactions along the $\tau$ direction, $r(\Delta l)$,
have to be determined
self-consistently from 
\begin{equation}
r(\Delta l) = {1 \over \sqrt{2 \pi}} 
\int_{-\infty}^\infty dz\,e^{-z^2/2}\, 
\langle \, S_{l_0} \, S_{l_0 + \Delta l} \,\rangle_{\overline{H}} \, ,
\label{r_sc_qu}
\end{equation}
where we can use any time slice for $l_0$ because of
translational invariance, and
$\langle \cdots
\rangle_{\overline{H}}$ indicates an average over the spins with weight
$e^{-\overline{H}}$. The order parameter $q$ is determined self-consistently
from 
\begin{equation}
q = {1 \over \sqrt{2 \pi}} 
\int_{-\infty}^\infty dz\,e^{-z^2/2}\, \langle \, S_{l_0}
\, \rangle^2_{\overline{H}}\, 
\, ,
\label{q_sc_qu}
\end{equation}
where again we can use any time slice for $l_0$ because of
because of translational invariance.

A breakdown of replica symmetry occurs when a divergence occurs in
the spin glass susceptibility $\chiSG$,
which is defined as follows.  If we change the field on site $i$ by
a small amount $\delta h_i$ then the expectation value of $\sigma^z_j$ changes
by an amount
\begin{equation}
\delta \langle \sigma^z_j \rangle = \chi_{ij}\, \delta h_i \, ,
\end{equation}
where the susceptibility $\chi_{ij}$ is given, according to linear response
theory, by
\begin{equation}
\chi_{ij} = \int_0^\beta d\tau \left[ \langle \sigma_j^z(\tau) \sigma_i^z(0) \rangle
- \langle \sigma_j^z\rangle \langle \sigma_i^z \rangle \right] \, ,
\end{equation}
where we note that single-spin expectation values are independent of $\tau$.
The spin glass susceptibility is then given by~\cite{class_qu}
\begin{equation}
\chiSG = {1 \over N} \sum_{i, j = 1}^N [ \chi_{ij}^2 ]\av \, .
\label{chiSG}
\end{equation}
Equivalently, $\chiSG$ gives the change in the spin glass order parameter $q$ when
the variance of the random longitudinal field is change by an amount 
$\Delta$ according to
\begin{equation}
\delta q = \chiSG\, \Delta \, ,
\end{equation}
which demonstrates that $\chiSG$ is the order parameter susceptibility for spin
glasses and the symmetry breaking field is the variance of the local
longitudinal fields. 

The expression for $\chiSG$ in terms of the parameters in the effective
one-dimensional model is given in Ref.~\cite{young:17}. One finds
\begin{equation}
\chiSG = {\chiSG^0 \over 1 - J^2 \chiSG^0},
\label{chiSGRPA}
\end{equation}
where
\begin{multline}
\chiSG^0 =  {1 \over \sqrt{2 \pi}}\, \int_{-\infty}^\infty
dz \, e^{-z^2/2}  \times \\
\left[ \sum_l \Delta \tau \Bigl( \langle \,
S_{l_0} S_{l_0+l}\, \rangle_{\overline{H}} - \langle \,S_{l_0}
\,\rangle^2_{\overline{H}} 
\Bigr) \right]^2 \, .
\label{lambda_r_qu}
\end{multline}
The denominator in Eq.~\eqref{chiSGRPA} is the
``replicon'' eigenvalue
$\lambda_r$ first calculated by AT~\cite{almeida:78} for the classical case.

It is straightforward to generalize these results to the case where we include
the 
Gaussian random longitudinal field of standard deviation $h$ in
Eq.~\eqref{ham}.  From
Eq.~\eqref{olineH} we see that $\overline{H}$ already has a 
term with a Gaussian random field of standard deviation $(\Delta \tau J)
q^{1/2}$, so it is sufficient to add another random field of standard
deviation $\Delta \tau h$. These two random fields can be combined into a single random
field of standard deviation $\Delta \tau [J^2 q + h^2]^{1/2}$ and so
$\overline{H}(z)$ in Eq.~\eqref{olineH} becomes
\begin{multline}
\overline{H}(z) = -(\Delta\tau J)^2 
\sum_{\langle l_1, l_2\rangle} \left\{\,r(|l_1 - l_2|) - q\, \right\}
S_{l_1} S_{l_2} \\
- K^\tau \sum_l S_l S_{l+1} -
\Delta \tau [J^2 q + h^2]^{1/2} z \sum_l S_l \, .
\label{olineH_h}
\end{multline}

%In this section we have used the discretized imaginary time formalism.
%Corresponding results for continuous imaginary time are described in
%Appendix~\ref{sec:continuous}.

\section{Numerical Results}
\label{sec:results}

\begin{figure}[htb]
\begin{center}
\includegraphics[width=0.9\columnwidth]{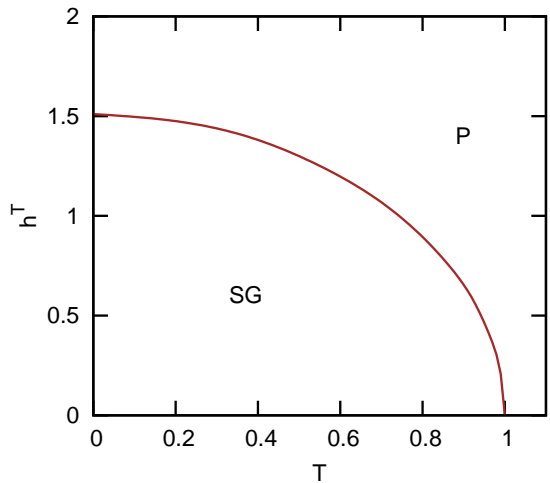}
\caption{\label{fig:Tc_hT} 
The phase diagram in the $h^T$-$T$ plane, determined from the vanishing of
the denominator of Eq.~\eqref{chiSGRPA} where
$\chiSG^0$ is given by
Eq.~\eqref{lambda_r_qu_q0}. The phases are spin glass (SG) and
paramagnet (P).
}
\end{center}
\end{figure}

In this section we shall mainly set $J = 1$.

\begin{figure}[htb]
\begin{center}
\includegraphics[width=\columnwidth]{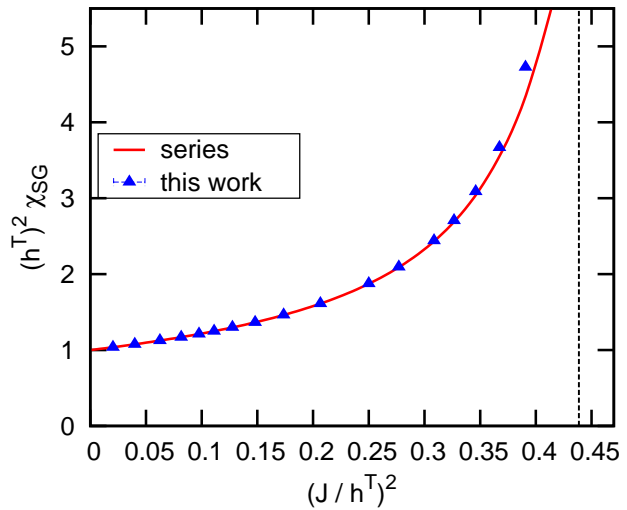}
\caption{\label{fig:chisg_T0} 
The triangles show results for $\chiSG$ from our calculations extrapolated
to $\Delta \tau = 0$ at very low
temperature ($T/J = 0.1$) in the paramagnetic phase.
The horizontal axis is $(J/h^T)^2$.
On the vertical axis we multiply $\chiSG$ by $(h^T)^2$ because, for $h^T \to \infty$,
perturbation theory gives $\chi_{SG} = 1 / (h^T)^2$.
The spin glass
susceptibility diverges at $ h^T_c/J \simeq 1.51$ as we shall see later, and
this is
indicated by the dashed vertical line. For the range of data shown there is a
gap in the spectrum so our results converge rapidly as $T \to 0$ and we find
that $T/J=0.1$ is low enough that the results are almost indistinguishable
from those at $T=0$. The line is the result of a 14-term
series expansion~\cite{singh:17} in
powers of $(J/h^T)^2$, evaluated for $T$ strictly equal to $0$. The agreement
is excellent until the critical point is approached where (a) we need to be
more careful in extrapolating our results to $T=0$ as we shall see later, and
(b) the solid line is the ``raw" series from Ref.~\cite{singh:17} so it does not show
a divergence at $h^T_c$.
}
\end{center}
\end{figure}

First we shall consider the case of 
no longitudinal field and zero spin glass order
parameter $q$ so we set $q=0$ in Eq.~\eqref{olineH}.
In addition, the expectation value in Eq.~\eqref{r_sc_qu} 
does not depend on $z$, so the $z$ integral trivially decouples and gives
one, and hence
\begin{equation}
r(\Delta l) = \langle S_{l_0} S_{l_0+\Delta l}\rangle \, .
\label{r_sc_qu_q0}
\end{equation}
Furthermore $\langle S(0)\rangle$ in Eq.~\eqref{lambda_r_qu} vanishes so
%the replicon eigenvalue is given by
\begin{equation}
\chiSG^0 = 
\left[ \Delta \tau \sum_l \langle \,
S_{l_0} S_{l_0+ l}\, \rangle_{\overline{H}} 
\right]^2 \, ,
\label{lambda_r_qu_q0}
\end{equation}
and we recall from Eq.~\eqref{chiSGRPA} that the critical point is when $J^2 \chiSG^0
= 1$. 

\begin{figure}[htb]
\begin{center}
\includegraphics[width=\columnwidth]{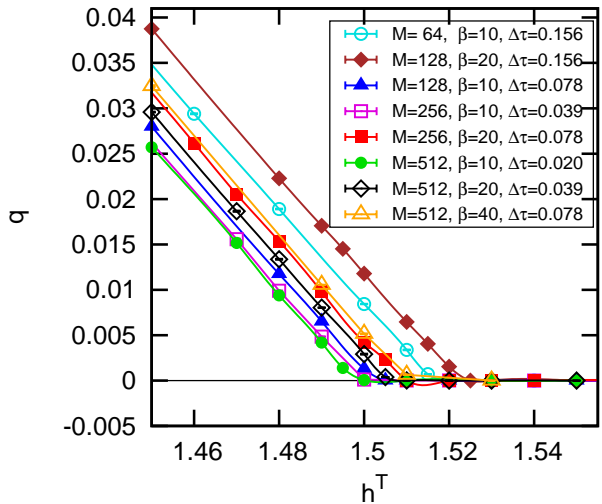}
\caption{\label{fig:q_enlarge} 
The spin glass order parameter $q$ near the quantum critical point at $T=0,
h^T = h^T_c$. The linear behavior shows that the order parameter exponent
$\beta$ has the value $\beta = 1$ in agreement with analytic
work~\cite{read:95}. 
}
\end{center}
\end{figure}

We perform standard Metropolis, single spin-flip Monte Carlo on a
one-dimensional chain of $M$ sites with periodic boundary conditions with
long-range interactions as specified in Eq.~\eqref{olineH} (with $q = 0$ for now).
In addition to the value of $M$ we also need to specify
the time-slice width $\Delta \tau$. We have to determine self-consistently the
$M/2$
parameters $r(\Delta l)$ where $\Delta l = 1, 2, \cdots M/2$. We start by
making a guess for the $r(\Delta l)$ and then run a high-precision Monte Carlo
simulation for these parameters to get the expectation values
$\langle S_l S_{l+\Delta l}\rangle$. Following Eq.~\eqref{r_sc_qu_q0} these
values are used for the next estimate of $r(\Delta l)$ and we iterate 
until convergence.  We did two hundred iterations, except for the small values
of $M$ where one hundred iterations were performed, and verified that
estimates of $\chiSG$ had converged at a much smaller 
number of iterations. We averaged over the
last quarter of the iterations and performed eight runs to estimate error bars
and improve statistics.
For $M \le 24 $ we were able to
use exact enumeration as well as Monte Carlo, which has the advantage of
there being no statistical errors. The exact enumeration results
served as a useful check on the Monte Carlo code.

The phase boundary in the $h^T$-$T$ plane is where the denominator in
Eq.~\eqref{chiSGRPA} vanishes and our results for this
are shown in Fig.~\ref{fig:Tc_hT}.
In the limit of $T \to 0$ we find the critical value of $h^T$ to be
\begin{equation}
\label{hTc}
h^T_c = 1.51 \pm 0.01 \, ,
\end{equation}
in agreement with earlier work of Yamamoto and Ishii~\cite{yamamoto:87}
who obtained $h^T_c = 1.506$ from a perturbation expansion. We will discuss
the error bar quoted
in Eq.~\eqref{hTc} in the context of Fig.~\ref{fig:q_enlarge} below.

As a check on our calculations
we compare our results in the low temperature limit in the paramagnetic phase with a recent
series expansion~\cite{singh:17}.
The leading correction due to a finite value of $\Delta \tau$ in Quantum Monte
Carlo
simulations varies as~\cite{fye:86}
$(\Delta \tau)^2$, so
we extrapolated our results for several values of $\Delta \tau$
to $\Delta \tau = 0$ by doing a linear fit in $(\Delta \tau)^2$. 
Our results at $T=0.1$, extrapolated to $\Delta \tau = 0$,
are in excellent agreement with the series results as shown
in Fig.~\ref{fig:chisg_T0}.

To investigate the stability of the RS solution we have to work out this
solution in the spin glass phase shown in Fig.~\ref{fig:Tc_hT}. Even with no
external longitudinal field, the presence of a non-zero spin glass order
parameter $q$ requires us to do the $z$ integral in Eqs.~\eqref{r_sc_qu},
\eqref{q_sc_qu}
and \eqref{lambda_r_qu},
as well as determine $q$ self-consistently. This is in addition to the
self-consistent determination of
$r(\Delta l)$ which we had before in the
paramagnetic phase. Doing the $z$ integral is, in general,
quite challenging so we will only get results  for small $q$ which means we stay
close to the phase boundary in
Fig.~\ref{fig:Tc_hT} (and eventually apply only a small longitudinal field).

We perform the $z$-integrals by Gauss-Hermite integration~\cite{press:92}
according to the following formula
\begin{equation}
\int_{-\infty}^\infty e^{-x^2} \, f(x) \, d x = \sum_{j=1}^L w_j f(x_j) \, ,
\end{equation}
where the weights $w_j$ and abscissas $x_j$ are tabulated and can also be
evaluated from scratch~\cite{press:92}. (We used tables available on the internet.) 
Since the integrands are even functions of $z$ we only need to take positive
values of $x_i$. To check for convergence we did calculations with 3, 6, 12
and 24 positive values.

In all cases we found that $\chiSG$ went negative on crossing the phase
boundary, indicating that RSB occurs \textit{everywhere} in the spin glass
phase in Fig.~\ref{fig:Tc_hT}. We present here only our results at low $T$,
where there has been the greatest
controversy~\cite{ray:89,mukherjee:15,mukherjee:17,goldschmidt:90,buttner:90,read:95},
as mentioned above.

\begin{figure}[htb]
\begin{center}
\includegraphics[width=\columnwidth]{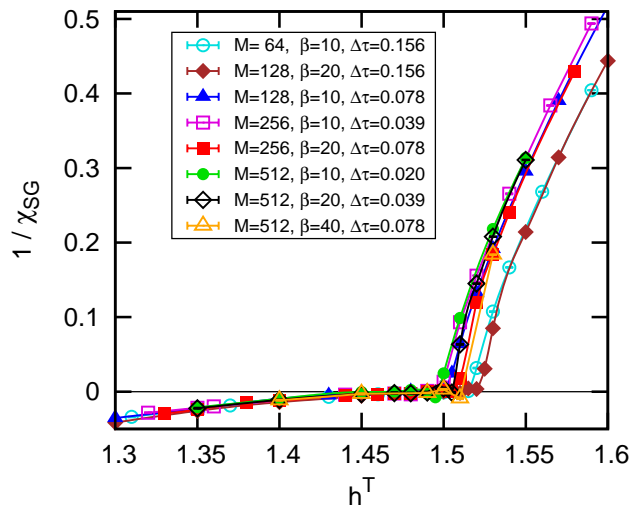}
\caption{\label{fig:chisg_instab} 
A plot of $1/\chiSG$, the inverse of the spin glass susceptibility, near
the quantum critical point (QCP) at $T = 0, h^T_c \simeq 1.51$. One sees
that $1/\chiSG$ tends to zero as the QCP is approached from above. Below
the QCP the spin glass order parameter becomes non-zero as shown in
Fig.~\ref{fig:q_enlarge}. Nonetheless, 
$\chiSG$ is negative for $h^T < h^T_c$. However this is impossible since $\chi_{SG}$
is a positive quantity, and hence 
the assumption of replica symmetry, made in the calculation, must be wrong.
}
\end{center}
\end{figure}

\begin{figure}[htb]
\begin{center}
\includegraphics[width=\columnwidth]{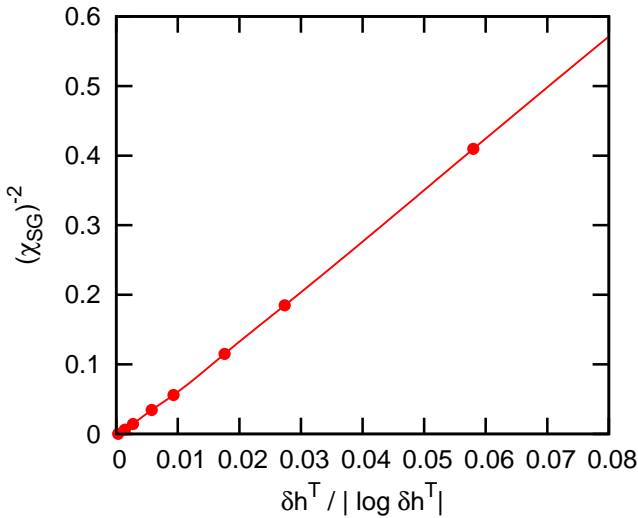}
\caption{\label{fig:chisg_sq} 
A plot of $1/\chiSG^2$ against $\delta h^T / |\log \delta h^T |$ for $N =
256, \beta = 20$, where $\delta h^T = h^T - h^T_c$ in which $h^T_c$, the critical
value, was taken to
be $1.508$. According to analytic predictions~\cite{ye:93,huse:93}, $\chiSG
\propto \left(|\,\ln \delta h^T\, | / \delta h^T\right)^{1/2}$, so the plot should be a straight line.
Clearly this form works very well. 
}
\end{center}
\end{figure}

Firstly we show our results for the spin glass order parameter $q$ near the
quantum critical point at $T=0,
h^T = h^T_c$. It clearly vanishes linearly showing that the order parameter
exponent $\beta$, defined by $q \propto (h^T_c - h^T)^\beta$, has the value 
\begin{equation}
\beta =1,
\label{beta}
\end{equation}
in agreement with analytic 
work~\cite{read:95}. Note that there is no finite-size rounding in this
approach since the effective one-dimensional model we simulate is a representation of the
original SK model in the thermodynamic limit. There are clearly small
corrections coming from the temperature
$T$ and the time slice width $\Delta \tau$ being not
precisely zero. Based on the scatter of the data in Fig.~\ref{fig:q_enlarge}
we estimate the value of $h_c^T$ to be $1.51 \pm 0.01$, as indicated in
Eq.~\eqref{hTc}.

\begin{figure}
\begin{center}
\includegraphics[width=\columnwidth]{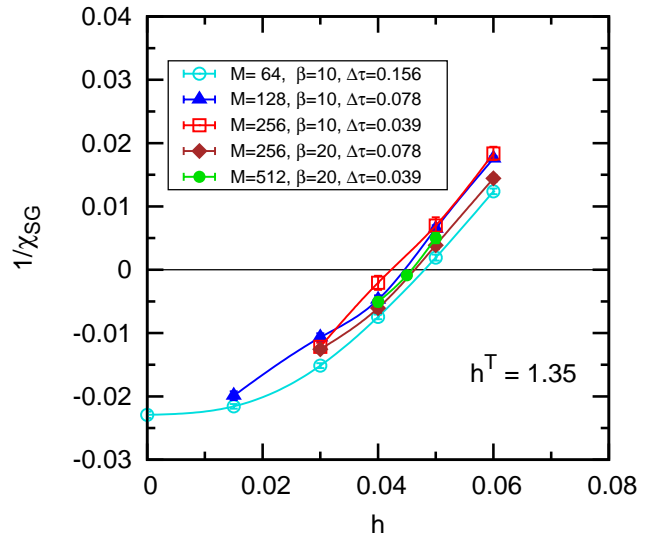}
\caption{\label{fig:chisg_h} 
A plot of $1/\chiSG$ against $h$ for $h^T = 1.35$ and very low temperature.
For $h \lesssim 0.045$,
the calculated $\chiSG$ is negative indicating that the RS solution is
unstable, whereas for $h \gtrsim 0.045, \chiSG$ is positive indicating the
RS solution is stable. The set of parameters $T=0, h=0.045, h^T = 1.35 $ lies
on the Quantum AT (QuAT) line, see Fig.~\ref{fig:hrAT} 
}
\end{center}
\end{figure}

\begin{figure}
\begin{center}
\includegraphics[width=\columnwidth]{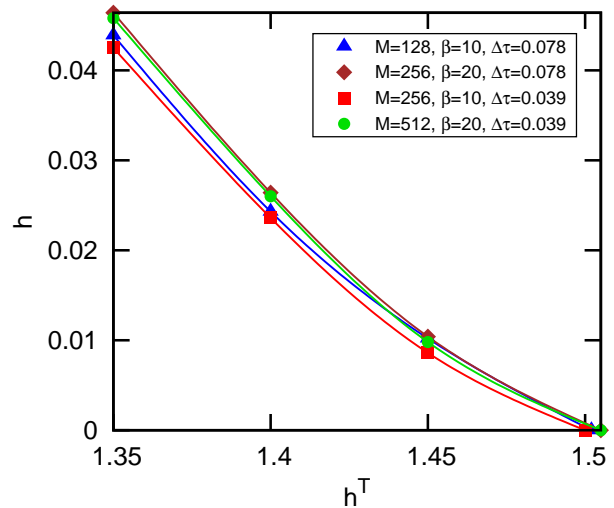}
\caption{\label{fig:hrAT} 
A plot of the QuAT line near the quantum critical point. 
}
\end{center}
\end{figure}

To investigate the stability of the RS solution (which has been assumed in our
calculations) we compute $\chiSG$ in the quantum spin glass phase and show
results for this quantity near the quantum critical point (QCP) in Fig.~\ref{fig:chisg_instab}. 
One sees that the computed $\chiSG$ is negative for $h^T < h^T_c$ with very
little dependence in this region on the precise values of $\Delta \tau $ and
$T$. However $\chiSG$ cannot really be negative so the assumption of replica
symmetry, made in deriving the expression for $\chiSG$, must be false. We
conclude that replica symmetry must be
broken in the quantum spin glass phase even down to $T = 0$. In the original
calculation of AT~\cite{almeida:78} for the classical spin glass, the negative
eigenvalue (the denominator in Eq.~\eqref{chiSGRPA} which is essentially $1/\chiSG$
near the instability) varies quadratically in the unstable region
near the classical critical point. From the
data in Fig.~\ref{fig:chisg_instab} it is plausible that the same
quadratic variation occurs near the QCP for $h^T < h^T_c$, though we cannot
determine the power with much accuracy, mainly because of the uncertainty in
the precise value of $h_c^T$.

According to analytical work~\cite{ye:93,huse:93} the spin glass
susceptibility should vary as
\begin{equation}
\chiSG \propto \left({|\ln \delta h^T | \over \delta h^T}\right)^{1/2} \, ,
\end{equation}
approaching the quantum critical point from the paramagnetic phase, 
where $\delta h^T = h^T - h^T_c$. Our numerical data fits this very well as
shown in Fig.~\ref{fig:chisg_sq}.

Including a longitudinal field $h$ we mapped out $1/\chiSG$ against $h$ for
different values of $h^T$ at low $T$, see Fig.~\ref{fig:chisg_h} for an example.
We then determined where $1/\chiSG = 0$ for different values of $h^T$ near $h^T_c$
and so were able map out the $T=0$ QuAT line near the vicinity of the QCP. The
results, which are shown in Fig.~\ref{fig:hrAT}, are largely independent of the values of
$T$ and $\Delta \tau$ shown.
The AT line for the classical spin glass
has the form $\delta T \propto h^{2/\phi}$ with $\phi = 3$. In the quantum
case, writing $\delta h^T \propto h^{2/\phi}$, the data in
Fig.~\ref{fig:hrAT} indicates that $\phi > 2$ but the
precision of the numerics does not allow us to determine the exponent
precisely.  We note that Ref.~\cite{read:95} finds analytically that same exponent $\phi =
3$ occurs everywhere including $T=0$. 

\begin{figure}
\begin{center}
\includegraphics[width=0.8\columnwidth]{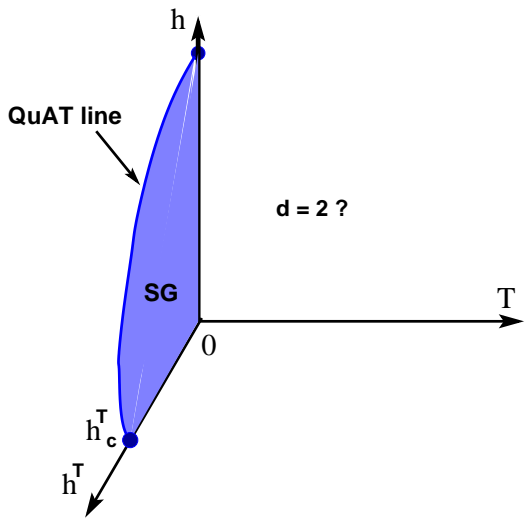}
\caption{\label{fig:quatline_2d} 
A speculative phase diagram for two dimensions. There is no spin glass phase
at finite-T but there is quantum spin glass phase at $T=0$ up to $h^T_c$, the critical
value of $h^T$, so there might possibly be a QuAT line in the $h$-$h^T$ plane
at $T=0$. While perhaps unlikely, one can not rule
out this scenario at present.
}
\end{center}
\end{figure}

Finally we note the possibility of a QuAT line even in two dimensions where
there is no spin glass phase at finite temperature, see
Fig.~\ref{fig:quatline_2d}.

\section{Conclusions}
\label{sec:conclusions}

We have shown that the replica symmetric solution of the quantum
Sherrington-Kirkpatrick model is unstable \textit{everywhere}
below a surface in the $h$-$h^T$-$T$ parameter space sketched in
Fig.~\ref{fig:quatline}, in agreement with
Refs.~\cite{read:95,goldschmidt:90,buttner:90}.
We suspect that the contrary results obtained
numerically in Refs.~\cite{ray:89,mukherjee:15,mukherjee:17} is due to inadequate treatment
of finite-size effects. In particular, their claim that there is a finite-temperature
multi-critical point is probably a mis-interpretation of crossover effects from the
\textit{zero}-temperature quantum critical point. The replica
symmetric region found in Ref.~\cite{thirumalai:89} is presumably due to the
inadequacies of the static approximation used in that paper.
We also note that Yao et
al.~\cite{yao:16}, who have performed experiments on a realization of the
quantum SK model with sizes $N \le 16$, find a fast ``scrambling'' once the
spins freeze, which seems to contradict our claim that replica symmetry is
broken everywhere in the spin glass phase. However, the sizes in the
experiments are very small and may not reflect the behavior in the
thermodynamic limit. 

In the method adopted here, the thermodynamic limit is
taken from the start.
In each of the three axis planes
there is a line of transitions where replica symmetry is broken.
In the $h$-$T$ plane this is the familiar de Almeida-Thouless (AT)
line. The phase boundary in the $h^T$-$T$ plane is plotted accurately in
Fig.~\ref{fig:Tc_hT}. We call the line of transitions in the $h$-$h^T$ plane
the quantum de Almeida-Thouless (QuAT) line.  If the QuAT line also occurs in
systems with short range interactions it may affect the performance of quantum
annealers~\cite{johnson:11} when solving spin glass problems in the presence
of a bias (i.e.~magnetic field).

\begin{acknowledgments}
I would like to thank Rajiv Singh for many stimulating and informative
discussions.
\end{acknowledgments}

\bibliography{refs,comments2}

\begin{thebibliography}{32}
\bibitem[{ran()}]{rand_hT}
\bibinfo{note}{One could also add disorder to the transverse field. We are not
  aware that this has been studied in detail, but expect that adding some
  additional disorder to an already strongly disordered Hamiltonian would not
  change the physics.}
\bibitem[{cla()}]{class_qu}
\bibinfo{note}{In the classical limit, Eq.~\eqref{chiSG} gives $$ \chiSG = {1
  \over T^2}{1 \over N} \sum_{i, j=1}^N [ \left(\langle \sigma^z_i
  \sigma^z_j\rangle - \langle \sigma^z_i \rangle \langle
  \sigma^z_j\rangle\right)^2 ]\av. $$ It is usual, in classical spin glasses,
  to omit the factor of $1/T^2$~\cite{binder:86} since it varies smoothly in
  the vicinity of the finite-$T$ transition, but here we are interested in the
  behavior of transitions as $T \to 0$, so it is essential to include it.}
\end{thebibliography}

\end{document}